# High-dimensional broadband non-Abelian holonomy in silicon nitride photonics

Youlv Chen[1], Xuhan Guo [1✉], Xulin Zhang [2✉], and Yikai Su [1✉]

1. State Key Laboratory of Advanced Optical Communication Systems and Networks, Department of Electronic Engineering, Shanghai Jiao Tong University, China
2. State Key Laboratory of Integrated Optoelectronics, College of Electronic Science and Engineering, Jilin University, Changchun, China
    ✉E-mail:guoxuhan@sjtu.edu.cn; xulin_zhang@jlu.edu.cn; yikaisu@sjtu.edu.cn

Non-Abelian geometry phase has attracted significant attention for the robust holonomic unitary behavior exhibited, which arises from the degenerate subspace evolving along a trajectory in Hilbert space. It has been regarded as a promising approach for implementing topologically protected quantum computation and logic manipulation. However, due to the challenges associated with high-dimensional parameters manipulation, this matrix-valued geometry phase has not been realized on silicon integrated photonic platform, which is CMOS compatible and regarded as the most promising flatform for next-generation functional devices. Here, we demonstrate the first non-Abelian holonomic high-dimensional unitary matrices on multilayer silicon nitride integrated platform. By leveraging the advantage of integrated platform and geometry phase, ultracompact footprint, highest order (up to six) and broadband operation (larger than 100nm) non-Abelian holonomy unitary matrices are experimentally realized. Our work paves the way for versatile non-Abelian optical computing devices in integrated photonics.

**Introduction**

Holonomy, which refers to the scenario that a quantum system evolves around a closed loop in a parameter space, is dramatically attracted for the Abelian and non-Abelian geometry phase factors acquired. In contrast to dynamic phase, geometry phase cannot be removed under gauge transformation since it is solely dependent on the evolution path in Hilbert space. One well-known geometry phase is Pancharatnam-Berry phase[1], which belongs to the Abelian (commutative) U(1) group transformation. Additionally, geometry phase can also represent high-dimensional unitary group U(m), which is non-Abelian in nature and can be generated by holonomic adiabatic evolution in degenerate subspace, as proposed by F. Wilczek and A. Zee in 1984[2]. This matrix-valued phase factor from non-Abelian gauge field is regarded as a promising route for various applications such as quantum computation[6].

So far, non-Abelian geometry phase has been proposed in various physical systems including cold atom systems[3], superconduct circuits[4], acoustic systems[15] and photonic systems[5-10]. Since the non-Abelian holonomic process can generate arbitrary-dimensional geometry phase matrix that is topologically protected and wavelength stable, it shows strong potential for manipulating photons on photonic chips as the obtained unitary matrix can be employed to design robust and versatile on-chip photonic devices working in broadband. Although a few proof-of-concept devices[7,8,10] have been experimentally realized by using the femtosecond laser direct writing techniques, the device suffers from larger footprint and thus the construction of higher-dimensional holonomic unitary matrix is difficult, which hinders the realization of more versatile non-Abelian operations and corresponding devices. Therefore, configuring the non-Abelian schemes in application-abundant and general fabrication technology-compatible photonic platforms and developing concomitant device architectures, are the next step.

Here, we report the experimental realization of ultra-compact on-chip high-dimensional non-Abelian holonomic computation on two-layer silicon-nitride-on-insulator (SNOI) photonic integrated platforms. In our design, the basic holonomy U(2) is achieved by effectively engineering the adiabatic evolution of coupling coefficients between four curved silicon nitride waveguides placed in two layers. By cascading and expanding this building block, we have successfully demonstrated various high-dimensional unitary matrices including diagonally dominant matrices, counter-diagonal dominant matrices and even the limiting case−non-Abelian braiding matrices, showing the excellent capability of our scheme in constructing arbitrary unitary matrices. In experiment, we have achieved high-dimensional matrices up to U(4) as well as arbitrary M×N matrices. Different from the unitary matrices achieved by using the dynamic phase effect such as that in Mach-Zehnder interferometers and directional couplers where the phase is highly sensitive to wavelength, matrix-valued geometry phase possesses unique broadband advantage because the non-Abelian holonomic process is insensitive to wavelength. Therefore, we have proposed an integrated platform scheme for non-Abelian holonomy with the advantage of ultracompact footprint (~600μm, 2 orders of magnitude shorter than reported), large bandwidth operation (larger than 100 nm) and highest matrix dimension (up to 6 order). It provides a feasible scheme for realizing ultra-large scale optical computing cascade mesh.

**Results**

Holomonic U(m) requires the construction of a m-dimension degenerate space, so we consider a Hamiltonian satisfying the chiral symmetry as follows :

$$\hat{H}(t) = \sum_{j=1}^{M+m} \sum_{i=1}^{M} \kappa_{i,j+M}(t) \hat{a}_i \hat{a}_{j+M}^+ + \kappa_{i,j+M}^+(t) \hat{a}_{j+M} \hat{a}_i^+ \tag{1}$$

This system consists of two groups, one with M particles and another with (M+m) particles. κ(t) represents the time-dependent hopping between particles from different groups, $\hat{a}_i$, $\hat{a}_i^{\mp}$ represents creation and annihilation operator. The diagonal elements of H(t) represent zero energy level of these identical particles. Chirality symmetry protects m-dimension degenerate states[15] and therefore this (2M+m)-particle system can be used to construct a U(m) holonomy. We start with the basic case, M=1 and m=2 to construct U(2) group, in which $\hat{H}$ is a four-order Hamiltonian and $\kappa$ is a 3 × 1 vector. This Hamiltonian can be reproduced by a four-waveguides system in two-layer integrated photonics platform. As shown in Fig. 1a, our scheme consists of two parallel straight waveguide A, B and two curved waveguide B, C. Waveguides B, X and C are on the first layer, while waveguide A is on the second layer. The material of waveguide is silicon nitride and the cladding material is silica. These four waveguides have identical cross-section to fulfill chirality symmetry. As the left inset of Fig. 1a shown, the width of waveguide is 800 nm and the height is 450 nm. Through coupled-mode theory[13-14], the light dynamic can be written as Schrödinger-like equation : H(z)|ψ(z)⟩=−i∂z|ψ(z)⟩. Where |ψ(z)⟩=[|w$_A$⟩,|w$_B$⟩,|w$_C$⟩,|w$_D$⟩] represents optical state in each waveguide. Evolution time t in Equation (1) is substituted by propagation length z. The diagonal element of Hamiltonian is the effective index of each waveguide, and κ=[κ$_{AX}$, κ$_{BX}$, κ$_{CX}$] describes the coupling coefficient between center waveguide X and adjacent waveguide A, B, C, which is determined by the gap between waveguide and wavelength (see Supplementary Fig. 1). The direct coupling between A, B and C can be negligible due to the large separation distance (Supplementary Fig. 1). κ$_{AX}$ is fixed in the whole evolution, κ$_{BX}$ and κ$_{CX}$ are modulated with varied g$_{AX}$, g$_{BX}$ by the curve path of waveguide A and B. In this arrangement, all the parameters of Hamiltonian can be attained, and its four eigenvalues as function of g$_{AX}$, g$_{BX}$ are depicted in Fig. 1b (at wavelength 1400nm). Obviously two of four eigenstates are degenerated (red sheet), and adiabatic cyclic loop of g$_{AX}$, g$_{BX}$ can accomplish holonomic evolution in degenerate subspace. Through this holonomy, the light injected from waveguide B or C can excite modes in degenerate subspace, and the initial and final state can be connected by a unitary matrix that $|\varphi_{final}\rangle = U|\varphi_{initial}\rangle$, where $|\varphi_{initial/final}\rangle = [|w_B\rangle, |w_C\rangle]$ (see Supplementary Note 1). U represents the path-order integral of Wilczek-Zee connection that $U = \hat{P} \oint \hat{A}$, where $A_{ij} = \langle D_i | \partial_\kappa | D_j \rangle$. It can also be expressed as $U = \exp(i\theta\sigma_y)$, where $\sigma_y$ is Pauli matrix, $\theta$ is the solid angle enclosed by cyclic loop in three-dimensin κ sphere (right inset of Fig. 1a, after normalization), so we can define the non-Abelian geometry phase $\theta$ as the U(2) characteristic angle. In principle, any target $\theta$ can be mapped to infinite paths in κ space. For the convenience of design and analysis, we propose a holonomy that evolves along the border of the first octant or approximately parallel to equator in κ sphere (red path in right inset of Fig. 1a), so that the target $\theta$ can be mapped to geometry space quickly. From this red path, it's clear that the loop can be divided into three steps. The start/final state is located at north pole of κ sphere (represented by the star position), with negligible $\kappa_{BX}, \kappa_{CX}$. For clockwise rotation, step 1 involves a three-waveguide stimulated Raman adiabatic passage (STIRAP), where waveguide B moves to X while C remains far from X; Then at step 2, waveguide C moves to X while B departs from X; Step 3 is also a three-waveguide STIRAP, which is the mirror symmetry of step 1. The Anticlockwise direction is the time reversal version of the process above. Therefore, the path of waveguide B and C determine the latitude of step 2 and target geometry phase $\theta$.

For a small $\theta$ which represents diagonally dominant U(2), we can have relative large $g_{AX}$, $g_{BX}$ at step 2. Specially, we take a unitary matrix with $\theta = \arctan(0.4)$ as an example ($\theta$ is integral of Wilczek-Zee connection, through the evolution of κ in Suppplementary Fig. 1d). Fig. 1c is the simulated light magnitude evolution in three steps through commercial software Lumerical FDTD[17]. Approximate 95% power remains in dark state[7] at the output port (waveguide B, C), indicating small adiabatic error in 600μm propogating length. On the contrary, for a large $\theta$ in represent of subdiagonal-dominate U(2), we should have small $g_{AX}$ and $g_{BX}$ to get small η in supplementary equation (2). However, the challenge is that the waveguides in close proximity will break the degeneration of modes, which will cause deviation of the target $\theta$. To suppress this, we utilize the void of PECVD to reduce some unwished coupling, increase the inter-layer spacing and so on (see Supplementary Note2). The experiment results fit with the mathematical derivation, as demonstrated in the subsequent chapter.

**Experimental result of U(2)**

The experiment setup is shown in Fig. 2a. A wideband laser from 1150nm to 1680nm is utilized as the input optical source. We use OSA to analyze the transmission spectrum of output light to detect the square of elements in unitary matrix. We select three typical U(2), $\theta > 45°$ (diagonally dominant matrix), $\theta \approx 45°$ (Hadamard matrix), $\theta < 45°$ (sub-diagonal-dominate matrix) to show the capacity of realizing arbitrary U(2). The specific values of the Wilson loop are approximate 1.82, 1.3, 1.07, respectively. Supplementary Fig. 2 illustrates the corresponding gap and κ, as well as integral of Wilczek-Zee connections. Fig. 2c-e depicts the calculated element of U(2) at different wavelength (the dash lines), and and the measured transmission spectrum of different inputs (the solid lines). We can observe the symmetry characteristic of the elements in diagonal and sub-diagonal ($|U_{11}|=|U_{22}|$, $|U_{12}|=|U_{21}|$), as well as broadband U(2) holonomy more than 100 nm The unique broadband property is attributed to the stable ratio of κ (φ,η in Supplementary Equation (2)) within a wide range of wavelength, despite the sensitivity of each κ to wavelength. To describe the accuracy, we define the Δθ (the difference between theoretical and measured θ), and the correlation (the average correlation of column vector between theory and experiment), which is shown in the insets of Fig. 2c-e. The measured magnitude of U(2) and the corresponding error bar at a wavelength of 1450nm is shown in the lower half of Fig. 2c-e.. More unitary experiment results can be found in supplementary material. These results indicates that these devices have relative high correlation (most >0.99), low angle error (most <6°) and low insert loss (<2dB).

**Scalability of integrated platform and experimental result of U(m)**

Leveraging the advantage of compact footprint of the integrated platform, these basic structures can be expanded to accomplish U(m) holonomy, which requires m-dimension degenerate sub-space. One famous structure is Reck -Zeilinger scheme[12] and the optimal proposal proposed by William R and Peter C et al[16]. Arbitrary high-order unitary can be realized by parallel and serial combinations of basic second-order unitary matrices. The optical depth is m and each basic U(2) can be calculated through mathematical deposition of target U(m). We refer to Hamitionian in equation (1), it means $M=\begin{cases} \frac{m}{2} \text{ or } \frac{m}{2}-1, m \text{ is even} \\ \frac{m-1}{2}, \quad m \text{ is odd} \end{cases}$. This (2M+m)-sites system construct m-dimension degenerate subspace. The integrated photonics structures of U(3) and U(4) are shown in Fig. 3a,b, the diagram below illustrates the simplified model of the optical mesh and mathematically deposition of U(2). We can see that U(3) is the cascaded of (2+3)-sites while U(4) is the alternatively cascaded of (2+4)-sites and (1+4)-sites. We measure the elements of U(3) and U(4) through the spectra of different inputs. These typical matrices are diagonally dominant and subdiagonal-dominate matrix, respectively. The measured magnitude and theoretical magnitude confirm well with each other, as shown in Fig. 3d,e. These scheme can be further expanded to arbitrary higher order M*M unitary matrix, and general M*N matrix can also be realized through SVG (singular value deposition ).

**Non-Abelian braid**

The above discussion pertains to general U(m), now we discuss the limiting case in which the characteristic angle $\theta=\frac{\pi}{2}$. This circumstance is the well-known non-Abelian braid. It requires the holonomy to evolves along the border of first octant in κ sphere (as shown in Fig. 1b). However, due to the fixed non-zero $\kappa_{AX}$ during evolution in Fig. 1a, characteristic angle θ can approach but can not reach the limiting value of $\frac{\pi}{2}$. In tackling this special situation, we change to the new scheme as shown in Fig. 4a. It can be viewed as the parallel transport of an isolated waveguide and a three-waveguide STIRAPs. The STIRAP consists of two outer straight waveguides and one center tilted waveguide, as depicted in Fig. 4b-c. It can support a zero mode that is degenerate with the mode in individual waveguide. The adiabatic evolution of this zero mode exhibits power transfer from one outer waveguide to another, while acquiring a geometry phase π. The simulation and experiment results are in supplementary Fig. 4-5, which also highlight the broadband characteristics. The parallel transport of STIRAP and an individual waveguide can be divided into three steps. At step 1, waveguide A,X and B form a vertical STIRAP, the titled waveguide X move toward B from A, $\kappa_{AX}$ decreases from its maximum value to negligible quantantility, while $\kappa_{BX}$ exhibits opposite trend. The light injected from B stimulating dark mode adiabatically transfers to waveguide A from B and acquires geometry phase π. At step 2, light in waveguide C experiences lateral STIRAP (B,X and C), transforms to waveguide B and acquires geometry phase π. Then two bends with same length were utilized to adjust the position of A and B, facilitating their convenient operation at step 3. At step 3, light in waveguide A goes through vertical STIRAP, transformed to waveguide C and acquired geometry phase π. So we can conclude the transformation as follows: $[\varphi_C, \varphi_B] \rightarrow [-\varphi_B, \varphi_C]$, which is non-Abelian braid. The corresponding $\kappa_{AX}$, $\kappa_{BX}$, $\kappa_{CX}$ evolution are concluded in Supplementary Fig. 3c, at least one null coupling coefficient ensures that the path evolves along the border of the first octant in the κ sphere (refer to Supplementary Fig. 3b).

The experimental result is shown in Fig.4a-b, the crosstalk is below 10 dB over a broadband of 100 nm. The realization of a two-order braid demonstrates the capability of the integrated photonics platform. Furthermore, this basic structure can be expanded to arbitrary higher-order braids through the optimal version of Reck-Zeilinger scheme. The traversal of all 5 types of three-order braid and 24 types of four-order braid mesh is shown in Supplementary Table. 1. The permutation of a three or higher order braid is not commutative, so we start with three-order braids with different permutations to illustrate this non-Abelian characteristic. The two kinds of generating operation of $B_3$ group are represented by $G_1$, $G_2^{8,15}$, where 
$G_1 = \begin{bmatrix} 0 & -1 & 0 \\ 1 & 0 & 0 \\ 0 & 0 & 1 \end{bmatrix}$, $G_2 = \begin{bmatrix} 1 & 0 & 0 \\ 0 & 0 & -1 \\ 0 & 1 & 0 \end{bmatrix}$ with basis [$|1\rangle, |2\rangle, |3\rangle$]. Fig. 5b shows the measured magnitude of different permutations of $G_1$ and $G_2$, and it is evident that $G_1G_2 \neq G_2G_1$. The experimental results of five and six-order braids, along with their corresponding simplified models are shown in Fig. 5c,d. The peaks (indicated by red bars) are consistent with theoretical prediction (output of simplified models), while the remaining portions (indicated by orange bars) represent the crosstalk. To demonstrate the broadband characteristic of our system, the transmission spectrum of five-order braid is shown in Supplementary Fig. 6. The input from $|2\rangle$ to $|5\rangle$ experience one braid, generally it's observed that crosstalk is around 20dB with 2.5dB insertion loss within wavelength range from 1300nm to 1500nm. It is worth noting that even when the input from $|1\rangle$ port experiences five-braid operation, crosstalk around 10 dB can still be maintained. This indicates the excellent expandability and broadband characteristic of the integrated platform.

**Conclusion**

In conclusion, we have realized high-dimensional non-Abelian holonomy on integrated photonics platform. Staring with basic U(2) group, high-order unitary matrix such as U(3) and U(4) have been realized through expanding of degenerate subspace. To tackle the limiting case, we also propose the parallel transport of STIRAP and individual waveguide to realize non-Abelian braids, up to six-order braid and over 100nm operation bandwidth has been observed. This topologically protected and robust high-dimensional geometry phase with ultra-compact footprint and large bandwidth will enable us to move towards large-scale integrated functional photonic devices of high-dimensional optical manipulation, such as the spatial optical switch, Thouless pump, large-scale matrices for optical neural networks. Furthermore, how to incorporate tunable non-Abelian holonomy to enhance the manipulation of light and photons is a valuable future field to explore. The proposed versatile integrated photonic platform may reveal more non-Abelian physics and lead to a new generation of on-chip devices for practical applications such as optical computing, optical routing and quantum computing.

**Method**

**Device simulation**

We use two kinds of simulation method for mutual confirmation, one is mathematical calculation of Wilczek-Zee connection integral through MATLAB, as the Equation (4) in Supplementary material, the other is 3D FDTD simulation for whole device. Coupling coefficient between two identical waveguides are calculated through Mode solutions that $\kappa = (n_s - n_a)/2$, where $n_s$, $n_a$ are effective index of symmetry and anti-symmetry supermodes, respectively. The mesh-order is set as 3 for trade-off between accuracy and simulation time. The refractive index of silicon nitride and silica are set as $n_{SiN}=2.03$ and $n_{SiO2}=1.47$ for better access to actual fabrication conditions.

**Device fabrication**

The fabrication process of two-layer silicon nitride device is compatible with CMOS technology. The devices are fabricated on a 500 μm silicon substrate wafer with a 3μm thermal-oxidation silica BOX layer.

For the first layer, a 450-nm-thick silicon nitride film is deposited by plasma-enhanced chemical vapor deposition (PECVD, Oxford). The designed patterns are defined by electronic beam lithography (EBL, Vistec EBPG-5200/JBX-9500FS), and fully etched through inductively coupled plasma reactive ion etching (ICP etching, NMC). After that, the chip is coated with silica as inter-layer dielectrics using PECVD.

For the second layer, similar processes are repeated as the first layer, such as depositing 450nm-thick-silicon nitride film, defining patterns through EBL, fully ICP etching, ect. Finally, a 3μm-thick silica cladding is deposited through PECVD.

After the fabrication, edge couplers need to be exposed for further experiments. Firstly, the double-sided alignment contact UV-lithography (SUSS MA6/MB6) is used to define patterns that protect devices while exposing the region outside the edge couplers. Then, ICP etching is employed to etch dielectrics until silicon substrate is exposed. After that, ICP deep-silicon etching (NMC) was carried out to etch the silicon substrate about 100μm. Finally, die sawing system (SPF-700) is used to slice the chip.

**Measurement setup**

Edge couplers are utilized to couple the light between single-mode fiber and silicon nitride waveguide, with the coupling loss is approximate 4dB per facet. The light is from a wideband source (Wuhan Leicheng Photonics Technology LTD). The light output is monitored by powermeter (JW3216) and the spectrum is analyzed through OSA (AQ6370C).

**Data availability**

The data that is relevant to this work is available from corresponding authors on reasonable request.

**Code availability**

The code that is relevant to this work is available from corresponding authors on reasonable request.

**Acknowledgements**

This work was financially supported by the National Key R&D Program of China (2021YFB2801903); Natural Science Foundation of China (NSFC) (62175151, 62105202, and 61835008). We also thank the Center for Advanced Electronic Materials and Devices (AEMD) of Shanghai Jiao Tong University (SJTU) and Tianjin H-chip Technology for the support in device fabrication, and Beijing MCF Technology LTD for chemical mechanical polishing (CMP) technology support. We would like to thank Mr. Shijun Qiao and Prof. Xincheng Ji for the helpful discussion in SiN fabrication process.

**Author contributions**

**Competing interests**

**Additional information**

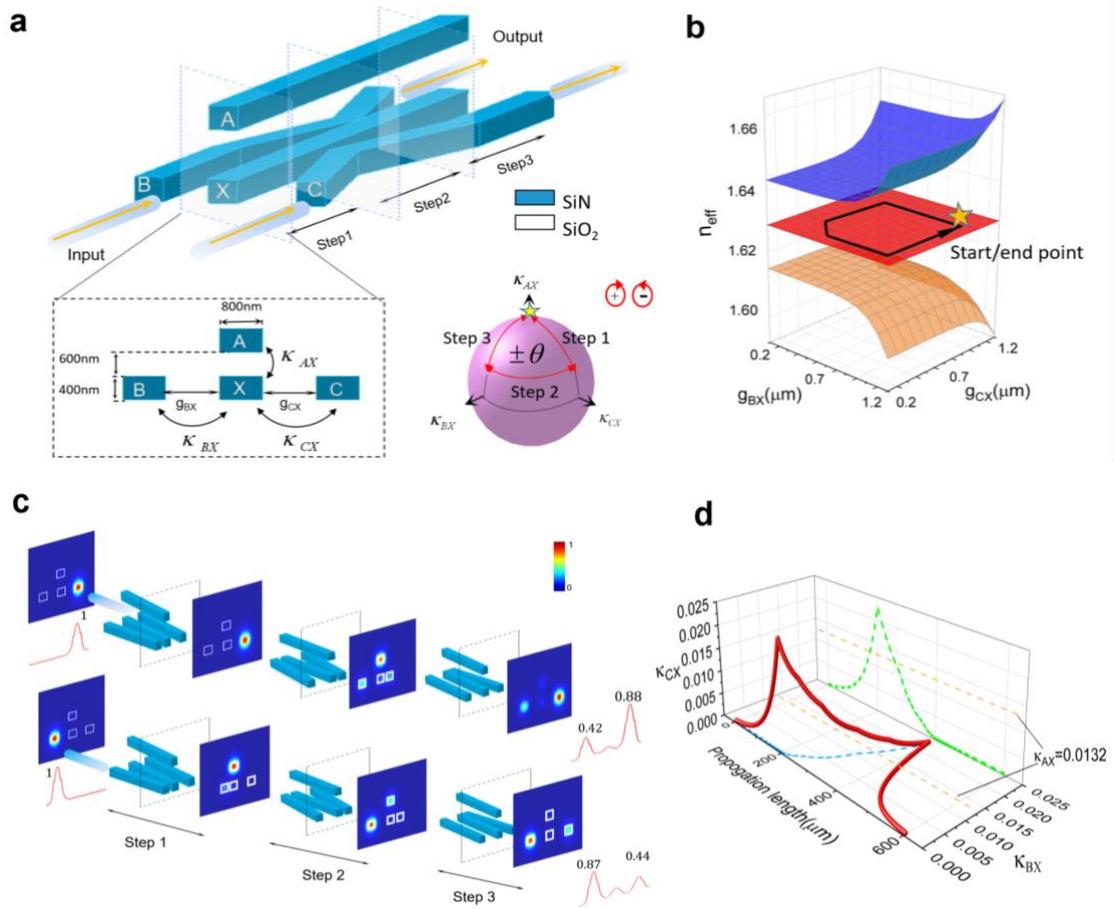

**Fig.1| Non-Abelian U(2) holonomy. a,** Schematic for four waveguides structure in two-layer SNOI platform. The left inset shows details of parameters in a crosssection, the right inset demostrate the holonmy evolve in $\kappa$ sphere (for better observation, coupling coefficients are all normalized as Supplementary Equation (3), which is equivalent to projecting them onto the spherical surface), the direction of rotation determines the sign of θ. **b,** Eigenvalue of the four-site Hamiltonian as function of $g_{AX}$, $g_{BX}$, the path shows the holonomy evolves in physical space. **c,** FDTD simulated light magnitude evolution with different input, the color represents the magnitude of different waveguides. **d,** Coupling coefficient evolution for unitary characteristic angle $\theta = \arctan(0.4)$ (mathematical calculated).

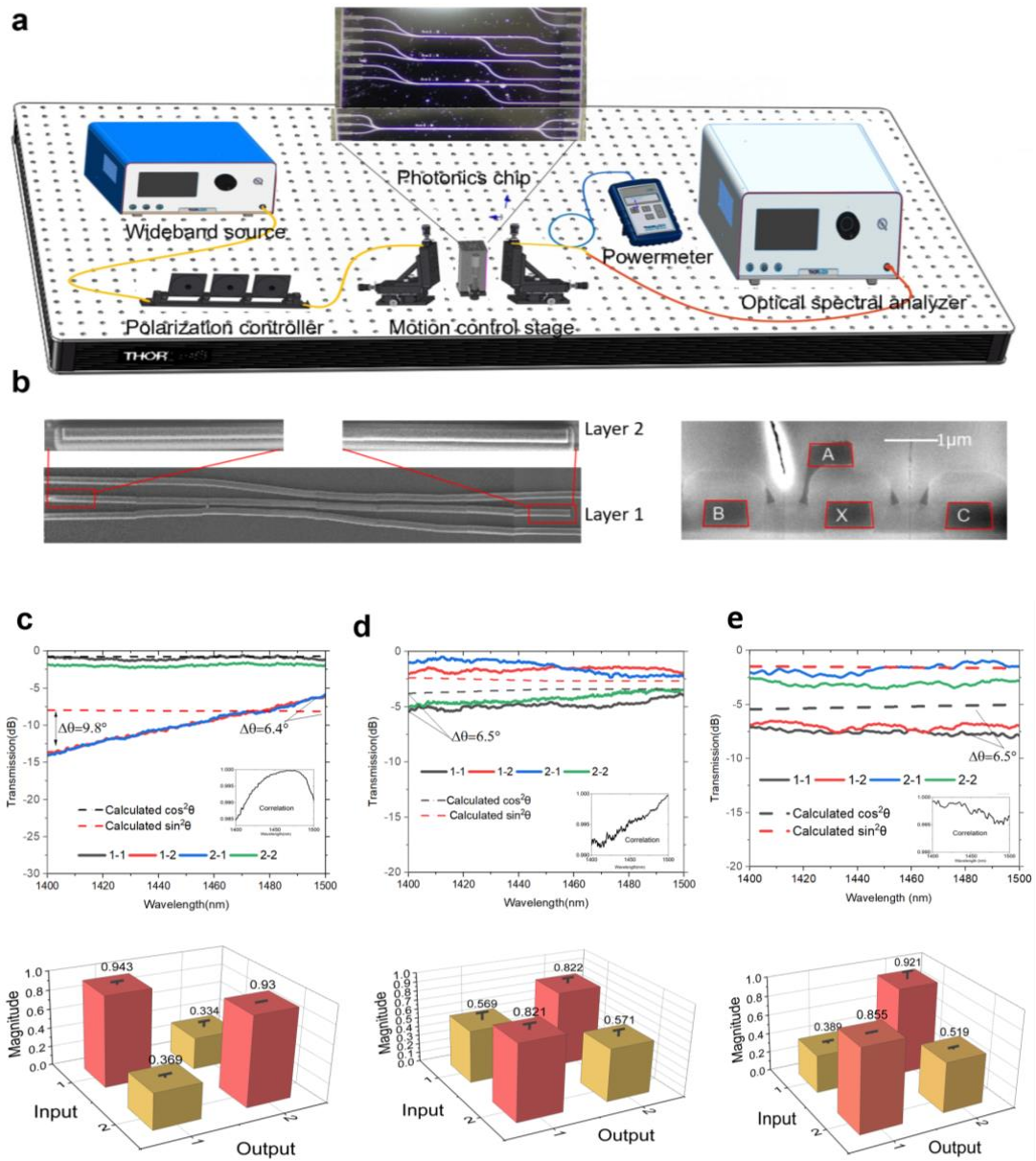

**Fig. 2| Experiment result for U(2) group. a,** Experiment setup and microscope of photonics chip. Light is supplied by wideband source, then it propogates through polarization controller and coupled in/out to photonics chip. The output light is monitor by powermeter and analyzed by optical spectrum analyzer. **b,** The left is SEM graph of layer 1 and layer 2, through splicing of a series SEM graph. The right is FIB image of cross-section at red dash line position. **c,d,e** Experiment result for three different U(2).

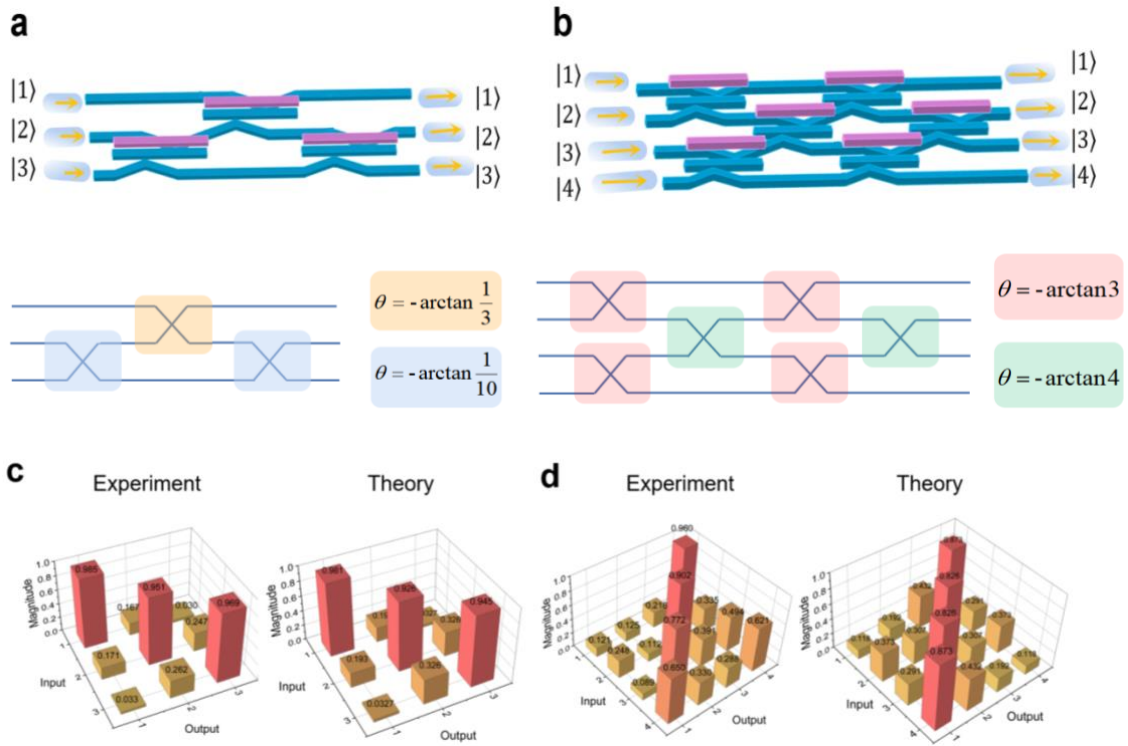

**Fig. 3| Experiment result for three-order unitary matirx and four-order unitary matrix. a,b,** Structure of U(3) and U(4) (up). The diagram below illustrates the simplified model of the optical mesh. The cross with distinct color represents different U(2). We can see that U(3) is the cascated of (2+3)-sites while U(4) is the alternatively cascated of (2+4)-sites and (1+4)-sites. **c,d,** Measured elements of U(3) and U(4) compared with theory.

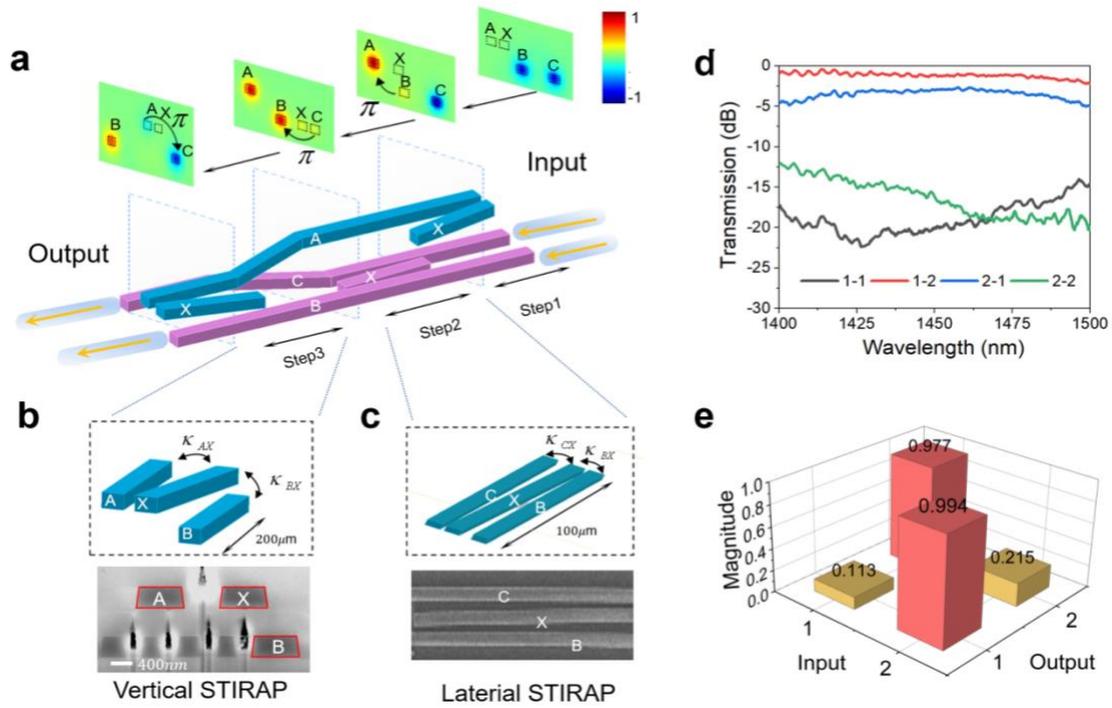

**Fig. 4| Structure and experimental results of Non-Abelian braid. a,** Structure diagram of non-Abelian braid in two-layer integrated photonics platform, the inset shows simulated light evolution in three steps (verticle-laterial-verticle STIRAP), evident power transfering and geometry phase acquired can be observed. **b,** Structure diagram of verticle STIRAP and its FIB graph of a crosssection. **c,** Structure diagram of laterial STIRAP and its SEM images. **d,** Experimental transmission spectra for non-Abelian braid. **e,** Measured output optical magnitude with different input at wavelength 1450nm after normalization.

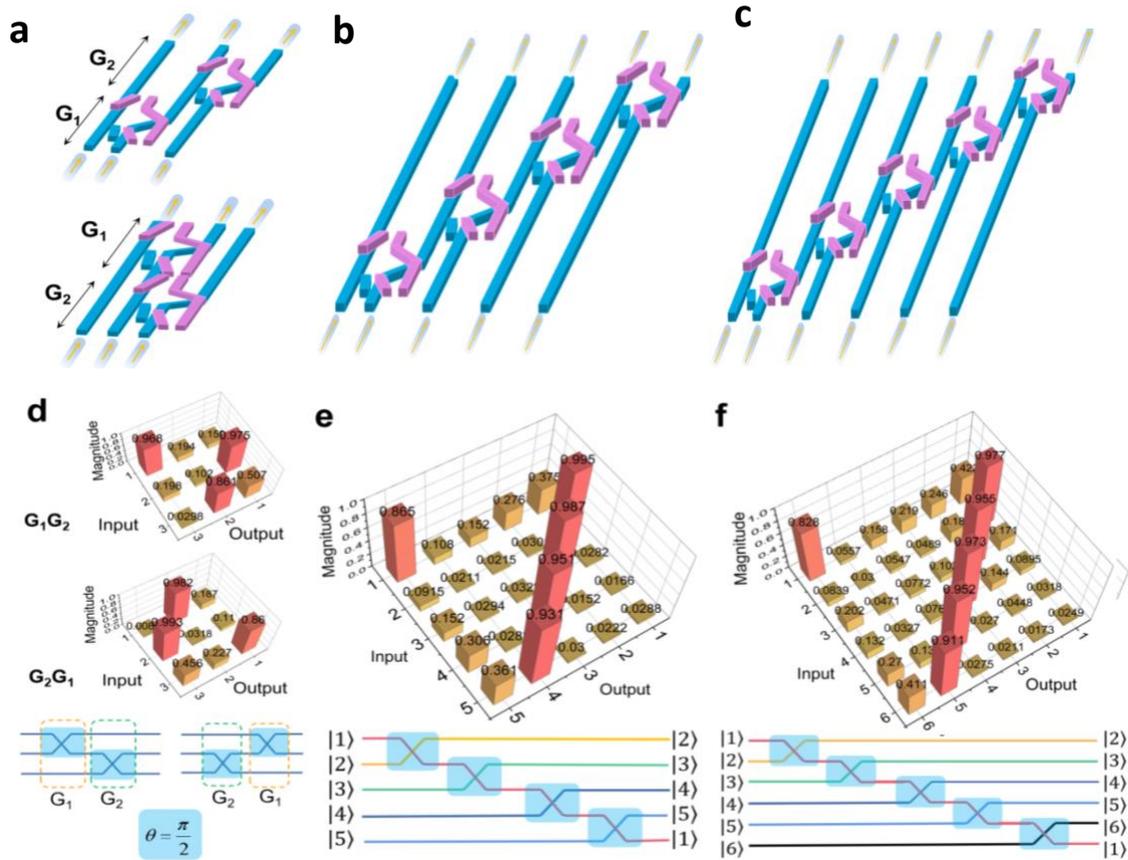

**Fig. 5| Experimental results of high-dimension Non-Abelian braid. a,** Structure diagram of five-dimension Non-Abelian braid in two-layer integrated photonics platform. **b,** Measured three-order Non-Abelian braid with different input at wavelength 1400nm after normalization. **c,** Measured five-order Non-Abelian braid with different input at wavelength 1400nm after normalization. **e,** Measured six-order Non-Abelian braid with different input at wavelength 1400nm after normalization.